\documentclass[twocolumn,nofootinbib]{revtex4-1}

\usepackage{amsmath,amssymb,graphicx,paralist,setspace,url,comment}

\renewcommand{\vec}[1]{\boldsymbol{\mathrm{#1}}}

\newcommand{\asterism}{\bigskip\par\noindent\parbox{\linewidth}{\small\centering{*}\\[-16pt]{*}\hskip 0.05em{*}}\par}%

\newcommand\myname{Viktor T. Toth}
\newcommand\myinit{VTT }
\newcommand\myurl{\url{https://www.vttoth.com/}}
\newcommand\ofours{of ours }

\begin{document}

\title{Feline gravity manipulation}

\author{\myname\\
\myurl}

\date{April 1, 2025}


\begin{abstract}
Since their domestication at the dawn of civilization, cats have been known for their uncanny ability to seemingly defy gravity. We conjecture that this innate ability of cats is real: uniquely in the animal kingdom, {\em felis catus}, possibly along with a few closely related species, are indeed capable of manipulating their passive gravitational mass. We explore this idea in the context of both general relativity and quantum physics. We reach the intriguing conclusion that a close study of the behavior of cats in a gravitational field might shed light not only on the mechanism of neutrino mass mixing but perhaps even on the most fundamental question in theoretical physics: a satisfactory unification of the theory of gravitation and quantum field theory.
\end{abstract}

\maketitle

\section{Introduction}

In 1922, the renowned British explorer Sir Archibald Whiskerton-Paws made a startling discovery in the long-lost city of Mau-nekhet-ra. Hidden deep in the tomb of the enigmatic Pharaoh Felinhotep III, Whiskerton-Paws uncovered a series of remarkably preserved murals depicting cats in flight. These images, dating back to the 20th Dynasty (circa 1100 BCE), indeed the reverence of cats throughout the history of ancient Egyptian society, puzzled Egyptologists for decades \cite{lyu2024}.

The uncanny abilities of cats have not escaped the attention of modern authors either. Of particular note is Heinlein's remarkable monograph \cite{heinlein1985cat}, which not only provides details on the gravitational capabilities of cats but also hints at a possible quantum mechanical connection, exhibited especially by younger felines.

\begin{figure}
\centering\includegraphics[scale=1]{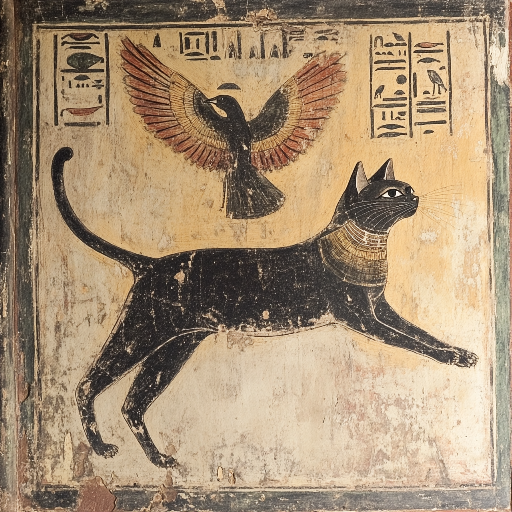}
\caption{\label{fig:egyptcat}One of the best preserved depictions of a flying cat from the burial chamber of Felinhotep III, circa 1100 BCE. [Image courtesy of {\em M. Journey}.]}
\end{figure}

Our contention, which we explore in this paper, is that these startling observations, evidenced across the ages in the form of both ancient Egyptian depictions (Fig.~\ref{fig:egyptcat}) and modern accounts, are not merely illusory. Cats indeed, uniquely in the animal kingdom, possess the ability to manipulate gravity.

More specifically, we conjecture that felines manipulate their effective gravitational mass between 0 and $2m$ where $m$ is their nominal mass.

To this end, we propose a theoretical model, which we tentatively call Modified Feline Gravitational Dynamics or MOFEGD. As we demonstrate below, it is possible to formulate MOFEGD in the context of general relativity, as a profound yet self-consistent modification of Einstein's theory of gravitation. Moreover, we find strong evidence that MOFEGD is a quantum theory. In addition to providing a mechanism that explains other aspects of startling feline behavior, our findings also provide a potential connection, thus insight, into the origin of the neutrino mass mixing mechanism.

We begin our study in Section~\ref{sec:background} by briefly reviewing the key relevant aspects of gravitation and the nature of inertial and gravitational mass. Next, we review available evidence of the observed dynamics of cat behavior and postulate a nonrelativistic, phenomenological model of feline-gravity interaction in Section~\ref{sec:dynamics}. Going beyond simple phenomenology, we propose a generally covariant modification of Einstein's theory of gravitation in Section~\ref{sec:GR}. Available evidence of the quantum nature of feline dynamics and a possible connection to neutrino mass mixing in the Standard Model are discussed in Section~\ref{sec:quantum}. We present our conclusions in Section~\ref{sec:summary}.

\section{Background}
\label{sec:background}

Our modern theories of gravitation, including general relativity and its Newtonian approximation in the limit of weak fields and low velocities, are based in part on the weak equivalence principle. The principle equates the inertia of a body, determined by its energy-content, with its effective ``gravitational charge'', its passive gravitational mass. As a key consequence, the body's mass cancels out in its equations of motion, leading to the conclusion that all bodies accelerate in a gravitational field at the same rate.

To see how this works, we can consider the context of Newtonian gravitation in conjunction with Newton's second law. The gravitational force $\vec{F}$ accelerating a body of mass $m$ in a gravitational potential $U$ is given by
\begin{align}
\vec{F}=-m\nabla U.
\label{eq:NewtonGrav}
\end{align}
Newton's second law, in turn, tells us that force is a product of mass and acceleration,
\begin{align}
\vec{F}=m\vec{a}.
\label{eq:Newton2}
\end{align}
Consequently, from (\ref{eq:NewtonGrav})--(\ref{eq:Newton2}), we have
\begin{align}
m\vec{a}=-m\nabla U\qquad\to\qquad \vec{a}=-\nabla U.
\end{align}
As we can see, the acceleration vector does not depend on the accelerating body's mass (or indeed, on any particular material property of the accelerating body).

These fundamental laws are tested at high accuracy in terrestrial laboratory experiments. We recognize of course that the Newtonian gravitational potential is just a useful approximation of the tensor field of gravitation in general relativity, but the principle remains the same: a test particle's mass cancels out even in the geodesic equations of motion.

On the other hand, ever since the 1930s there has been growing evidence \cite{Zwicky1933,Zwicky1937} that, on the scale of galaxies and beyond, Newton's (and therefore, of course, Einstein's) law of gravitation requires revision.

As a matter of fact, the model that is arguably the best known is not even relativistic. Mordechai Milgrom's Modified Newtonian Dynamics (MOND) \cite{Milgrom1983} is an {\em ad hoc} modification of Newton's second law:
\begin{align}
    \vec{F}=m\vec{a}\qquad\to\qquad\vec{F}=m\vec{a}\mu(a/a_0),
\end{align}
where $\mu(a/a_0)$ is an interpolating function that is characterized by the limits $\mu(a/a_0)\to 1$ for $a\gg a_0$, and $\mu(a/a_0)\to a_0/a$ for $a\lesssim a_0$, where $a_0\sim 1.2\times 10^{-10}~{\rm m}/{\rm s}^2$ is the MOND acceleration scale. MOND correctly reproduces the rotation and dynamics of spiral galaxies ``by design'', and it may be testable here in the solar system \cite{Toth2025a}. On the other hand, MOND lacks a robust generally covariant theoretical foundation, and it is challenged by cosmological data.

Several other modified theories of gravitation involve fields with a nonzero rest mass and consequently, finite range. At the Newtonian level of approximation, such theories are characterized by a Yukawa-type modification of Newton's gravitational potential \cite{Adelberger2003}. We represent the Yukawa-potential and its gradient (the acceleration) as follows:
\begin{align}
\Phi&{}=-\frac{GM_\odot}{r}\big[1+\alpha(1-e^{-\mu r})\big],\\
\vec{a}=\nabla\Phi&{}=-\frac{GM_\odot}{r^3}\big[1 + \alpha\big(1 - (1 + \mu r)e^{-\mu r}\big)\big]\vec{r},\label{eq:Yukawa}
\end{align}
with $\alpha$ and $\mu$ characterizing the strength and range of the Yukawa-interaction, respectively. This representation offers the advantage that in the $r\to 0$ limit it reduces to Newton's law of gravitation with $G$ being Newton's constant. Otherwise it is equivalent to the more popular $\Phi=-(GM_\odot/r)(1+\alpha e^{-\mu r})$, after the substitutions $G\Leftrightarrow (1+\alpha)G$, $\alpha\Leftrightarrow -\alpha/(1+\alpha)$. Yukawa gravity may also be testable in solar system experiments  \cite{Toth2025a}.

A common trait in MOND, Yukawa-gravity and many other proposed modifications of Newtonian dynamics, is the assumption that a body's passive gravitational mass (which determines how it responds to a gravitational field) and
inertial mass
are equal. In other words, Galileo's principle of gravitation---namely that the gravitational acceleration is independent of a falling body's mass or material composition  \cite{Galilei1638,Galilei1914}---is assumed valid. It is this principle that is put to the test by the studied behavior of feline dynamics.

\section{The motion of cats -- brief review}
\label{sec:dynamics}

How do cats move? Throughout human history, most civilizations made note of our feline companions' unique ability to move with grace and survive the seemingly unsurviveable, such as falls from great height. ``Cats always land on their feet,'' goes the common wisdom in the English language. Other languages offer similar expressions. Cat-like behavior is often associated with grace and agility.

Did cats acquire such reputation merely as a result of their unique physiology, an agile body, good reflexes, a brain that is especially adept at swift, precise movement? Or is there more to the observed behavior of cats?

These questions have fascinated generations of researchers, among them physicists, over the ages. And the interest did not wane. One recent study \cite{Studnicka2016} considered free-falling cats in the presence of air resistance. Although they even performed actual experiments, the use of an inanimate model (a ``plush toy'') raises questions about the validity of results: If cats indeed have unique abilities, how can we expect a simple mechanical simulacrum to exhibit similar behavior? Another, more recent study \cite{Challis2023} utilized an oversimplified model that considered only two fundamental forces (gravity and air resistance) when calculating a cat's motion through the air. We know that even unpowered (glider) aeroplanes cannot be modeled this simplistically, and powered flight requires the modeling of at least four forces (weight, lift, thrust and drag) even when the object -- be it an aeroplane or a cat -- is modeled as a point particle.

Needless to say, cats are not point particles and their interactions with their environment can be very complex. This is recognized in a yet more recent study \cite{Biasi2024} although the focus here was on cat-human interactions, with physics relegated to a secondary role.

\begin{figure}
\centering\includegraphics[scale=1]{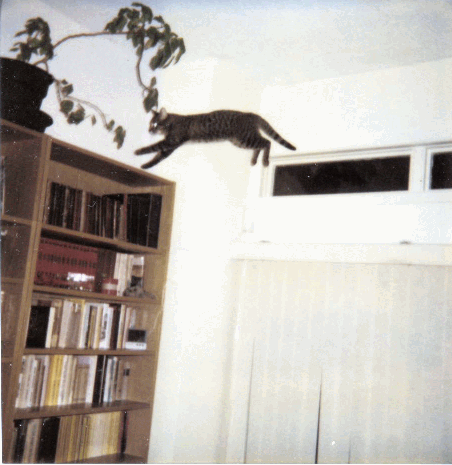}
\caption{\label{fig:mycat}Rare mid-flight image, taken using a Polaroid instant camera ca. 1999, of Marzipan, the author's former cat, exercising his ability to levitate by effectively reducing his passive gravitational mass to zero. [Image courtesy of {\em I. Czako.}]}
\end{figure}

\begin{figure}
\centering\includegraphics[scale=0.6666666667]{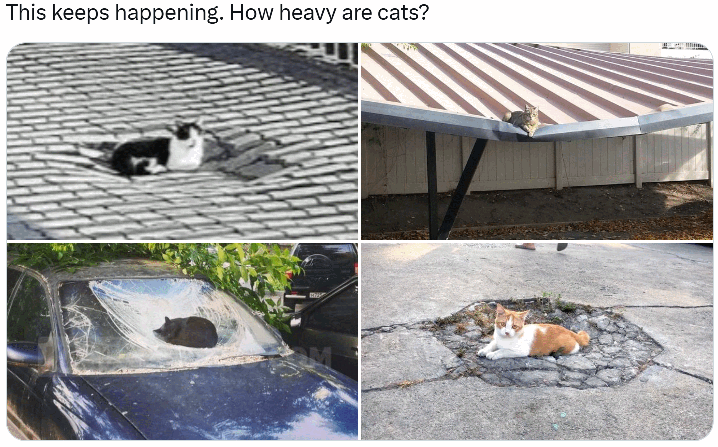}
\caption{\label{fig:heavycat}A popular Internet meme offering uncorroborated photographic evidence of cats appearing heavier than anticipated. From \protect\url{https://x.com/fuller_si/status/1388106214324662272}}
\end{figure}

Let us therefore focus on the physics of cat dynamics. Specifically, we begin with the assumption, or postulate, that cats have the ability---be it conscious or instinctive---to adjust their passive gravitational mass. Our postulate is based on the observed behavior of cats (See Figs.~\ref{fig:mycat} and \ref{fig:heavycat}.) While their passive gravitational mass is subject to manipulation, it appears that the inertial mass of felines remains constant. This is particularly evident when felines need to make a sharp turn on a smooth surface. Often, the feline slides several ten centimeters or more, trying to overcome inertia, even as their passive gravitational mass is very obviously reduced, considering their overall trajectories. As a matter of fact, the reduction of their passive gravitational mass may be a contributing factor: it also implies reduced traction, making it even harder for a cat to make a sharp turn.

This has profound implications, since this ability is a {\em bona fide} violation of the weak equivalence principle, notably the equivalence of inertial and passive gravitational mass. It may also explain the legendary feline ability to ``always land on their feet'', albeit it is not a subject of our current study.

Rather, we focus on the equivalence principle violating aspects of this feline ability. We do not discuss any possible biological or biochemical mechanism that may be behind this ability\footnote{While reasonably well-versed in the subjects of gravity, quantum field theory, or cats, I have no qualifications when it comes to molecular biology or biomechanics, which is the main reason why I leave these questions untouched. -- \myinit}. Our goal is to simply construct a sensible model, a viable extension of the general theory of relativity that would accommodate this unusual ability routinely exhibited by one of the most graceful species that exist in our biosphere.

In the nonrelativistic context, this ability to manipulate passive gravitational mass is easily modeled phenomenologically. In this representation, Newton's second law (\ref{eq:Newton2}) remains intact. However, when we split the total force acting on the cat as a sum of gravitational and nongravitational forces, we must introduce a modification of Newton's law of gravitation (\ref{eq:NewtonGrav}):
\begin{align}
\vec{F}=\vec{F}_{\tt nongrav} + \vec{F}_{\tt grav} = \vec{F}_{\tt nongrav} - \kappa(\psi)m_{\tt cat}\nabla U,\label{eq:MOFEGD}
\end{align}
where $m_{\tt cat}$ is the cat's nominal passive gravitational mass, while $0\le \kappa(\psi)\lesssim 2$ is a weight function that is determined by the cat's internal state $\psi$. What this expression asserts is that the internal state of the cat can result in a variation of the cat's effective gravitational mass between 0 and $\sim 2m_{\tt cat}$.

Expression (\ref{eq:MOFEGD}) is the phenomenological representation of our Modified Feline Gravitational Dynamics or MOFEGD, a theoretical framework representing the gravitational behavior of {\em felis catus}.

It is important to stress that $\kappa(\psi)$ modifies {\em only} the gravitational mass of the cat; the feline's inertial mass is unaffected. Therefore, the combination of Newton's gravitational law and Newton's second law must also be modified:
\begin{align}
m_{\tt cat}\vec{a}=-m_{\tt eff}\nabla U,
\end{align}
where $m_{\tt eff}=\kappa(\psi)m_{\tt cat}$ is the effective mass of the cat. Canceling out $m_{\tt cat}$, we are left with the equation,
\begin{align}
\vec{a}=-\kappa(\psi)\nabla U,
\end{align}
in clear violation of the weak equivalence principle. This is the essential statement of MOFEGD at the Newtonian level: $\kappa(\phi)$ serves as a gravitational charge-to-mass ratio, and we assert that cats can manipulate this quantity.

A modern gravitational theory of course cannot be limited to nonrelativistic phenomenology. Consequently, we now turn our attention towards general relativity, to see if we can incorporate a generally covariant representation of feline behavior in the framework of Einstein's theory.

\section{The general relativistic cat}
\label{sec:GR}

The generally covariant mechanism that we choose in order to introduce possible violations of the weak equivalence principle relies on postulating a scalar field in addition to the tensor field of Einstein gravity.

The archetypal scalar field theory is Jordan--Brans--Dicke theory \cite{Jordan1955,BD1962}. Although appealing, the theory has been effectively ruled out by precision solar system observations \cite{Will2000}. We nonetheless feel justified in introducing a scalar field theory, as our goal is not to seek a modification of Einstein's theory that still obeys the weak equivalence principle. Rather, we specifically use a scalar theory to introduce a {\em deviation} from the weak equivalence principle, to account for the observed unusual behavior of cats.

To wit, we envision a theory governed by the following generic action:
\begin{align}
S &{} = \frac{1}{16\pi G} \int d^4x \sqrt{-g} [R - 2\nabla_\mu\phi\nabla^\mu\phi - V(\phi)] \nonumber\\
&{} + S_{\tt cat}(\psi,\phi)+S_m,\label{eq:Lcat}
\end{align}
where $G$ is Newton's constant of gravitation, $g$ is the determinant of the spacetime metric $g_{\mu\nu}$, $R$ is the Ricci-scalar, $\phi$ is a scalar field, $V(\phi)$ is its self-interaction potential (and may be null), and $S_m$ is the matter action. The cat action, in turn, represented by $S_{\tt cat}(\psi,\phi)$, depends on $\psi$ that represents the cat's state, interacting with the scalar field $\phi$.

As usual, we expect that the variation,
\begin{align}
\frac{-2}{\sqrt{-g}}\frac{\delta (S_m+S_{\tt cat})}{\delta g_{\mu\nu}} = T_m^{\mu\nu}+T_{\tt cat}^{\mu\nu},
\end{align}
yields the stress-energy tensor of matter (including cats), while the variation,
\begin{align}
\frac{-1}{\sqrt{-g}}\frac{\delta S_{\tt cat}}{\delta\phi}=\mu,
\end{align}
yields a scalar charge that characterizes the feline-scalar coupling as a function of the coordinates. In the case of a single, point-like cat in state $\psi$ following a world line parameterized as $x_{\tt cat}^i(x^0)$, we expect
\begin{align}
\mu=\kappa(\psi)\delta^3(x^i-x_{\tt cat}^i(x^0)),
\end{align}
where $\delta^3$ represents the three-dimensional Dirac delta function.

Looking at the Lagrangian (\ref{eq:Lcat}), one may have concerns that the kinetic and potential terms characterizing the scalar field might conflict with existing observations constraining background cosmic energy density. We address these constraints by noting that
\begin{inparaenum}[a)]
\item the scalar field kinetic term can be arbitrarily small, and
\item the scalar field self-interaction potential, even if large, contributes a ``dark energy'' type equation of state, $w=p/\rho\simeq -1$, where $p$ and $\rho$ represent the corresponding magnitudes of pressure and density, respectively, modeling the self-energy contribution as an isotropic perfect fluid.
\end{inparaenum}
Indeed, is is even conceivable that this scalar field is responsible for the observed acceleration of the universe, its self-energy potential playing the role of dark energy.

There is another intriguing connection between feline gravitational behavior and general relativity. A recent paper \ofours \cite{Toth2024b} argues that rigorously constraining the metrical tensor of gravitation to be symmetric leads to a subtle but important modification of Einstein's field equations. Specifically, the stress-energy-momentum tensor $T_{\mu\nu}$ characterizing matter on the right-hand side is replaced by its explicitly symmetrized version, $T_{(\mu\nu)}$, whereas any nonsymmetric contribution to $T_{\mu\nu}$ remains completely unconstrained by gravity.

It is generally assumed that $T_{\mu\nu}$ is always symmetric for any ``sensible'' configuration of matter in classical field theory. We argue, however, that this need not be the case when cats are present: indeed, the frequently observed, ``uncanny'' twisting and turning behavior that cats often exhibit may be direct evidence of this. Meanwhile, it is well known that in the quantum theory, certain forms of matter, e.g., spinor fields, naturally lead to the presence of a nonzero spin contribution to the stress-energy tensor, in the form of an antisymmetric term. Customarily, these terms are canceled out using {\em ad hoc} modifications of $T_{\mu\nu}$ but what if it is unnecessary?

To this effect, we propose the following modification of Einstein's field equations in the presence of cats:
\begin{align}
R_{\mu\nu} - \tfrac{1}{2}Rg_{\mu\nu} + \Lambda g_{\mu\nu} = 8\pi G (T_{\mu\nu} + T_{(\mu\nu)}^{\tt{cat}}),
\end{align}
where, as usual, $R_{\mu\nu}$ is the Ricci-tensor of spacetime and $\Lambda$ is a cosmological constant, while $T_{\mu\nu}^{\tt cat}$ is a {\em nonsymmetric} stress-energy-momentum tensor characterizing cats. The gravitational field equations thus leave the antisymmetric part, $T_{[\mu\nu]}^{\tt cat}$, unconstrained. This part of the cat stress-energy-momentum tensor may best be described by a quantum theory of cats, which we discuss in the next section.

\section{Quantum cat theory}
\label{sec:quantum}

As we explored the classical gravitational behavior of cats in the context of both Newtonian and relativistic physics, we left open the key question: How do cats accomplish the seemingly impossible, manipulating their passive gravitational mass? To seek plausible answers, we now turn to the quantum theory. Specifically we draw, for inspiration, from the currently prevailing model of neutrino masses.

The puzzling discovery of a deficit in the observed density of electron neutrinos of solar origin, combined with the later realization that these neutrinos ``oscillate'' into a muon neutrino state en route, led to a daring conclusion. This behavior is now seen as evidence that neutrinos not only have masses, but that their mass and flavor eigenstates do not coincide: determining the (electron, muon, tau) flavor of a neutrino introduces a mass uncertainty and conversely, measuring the neutrino energy may change their observed flavor \cite{Burgess2007}. Phenomenologically, this behavior can be represented by supplementing the Lagrangian of the Standard Model with a mass mixing term:
\begin{align}
{\cal L}={\cal L}_{\tt SM}-\tfrac{1}{2}(m^{ij}\bar{\nu}_i P_L\nu_j+\text{c.c.}),
\end{align}
where $\nu_i$ is the $i$-th ($i=1..3$) neutrino field and $m^{ij}$ is an arbitrary $3\times 3$ symmetric matrix, the Pontecorvo--Maki--Nakagawa--Sagata (PMNS) mass mixing matrix. (The projection operator $P_L$ is included in this expression to account for neutrino handedness; c.c. stands as usual for complex conjugate.)

If $m^{ij}$ is diagonal, its three nonzero elements just represent the respective neutrino masses. However, if off-diagonal terms are present, they represent interactions between different neutrino species, which allow neutrinos to change flavor between source and detection. It is this mechanism that is believed to be at the root of the observed solar electron neutrino deficit and the corresponding observation of excess muon neutrinos.

Could something similar govern the quantum behavior of cats? To this end, let us entertain the idea of three cat states. In addition to the normal cat state, a cat may be in a ``fat cat'' state when its passive gravitational mass doubles, and a ``cheshire cat'' state in which its passive gravitational mass vanishes.

Clearly though, this cannot be the full story. Cats do not behave like neutrinos. As we noted earlier, the inertial mass of cats does not appear variable. This leads us to realize that a mass mixing term for cats must necessarily govern not the free cat field but the coupling between cats and the scalar field $\phi$ that we postulated earlier. Again representing the cat state using $\psi$, we therefore write down the interaction term,
\begin{align}
V(\psi,\phi)=-\phi(m^{ij}\hat\psi_i \psi_j+\text{c.c.}),
\end{align}
with $i=[C,S,F]$ representing the cheshire, standard, and fat cat states, and $m^{ij}$ serving as the feline mass mixing matrix, analogous to the PMNS neutrino mass mixing matrix. The theory is completed by including a term representing the cat's inertial mass (modeling the cat as a perfect fluid appears an especially suitable representation, as cats are known to exhibit perfect fluid behavior \cite{Fardin2014}) along with a suitable free field term for the scalar field $\phi$.

\begin{figure}
\centering\includegraphics[scale=1]{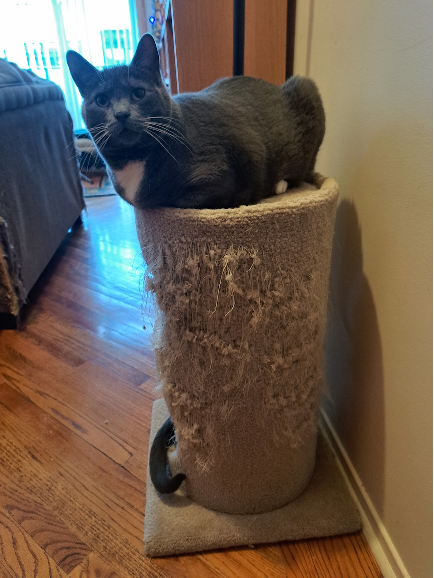}
\caption{\label{fig:quantumcat}Capturing the author's cat Rigby while quantum tunneling from the bottom to the top floor of a cat house, in early 2025. Note the tail. [Image courtesy of {\em I. Czako}.]}
\end{figure}

Another potential concern arises considering the nature of cats. Cats are obviously not elementary particles. How is it possible, then, to represent them using a mass mixing mechanism designed for quantum fields? We propose that altough in general, cats are non-coherent many-particle systems, they {\em are}, in fact, coherent with respect to species and passive gravitational mass. Therefore, with respect to the operators $\hat{\cal S}$ (species) or $\hat{\cal M}$ (passive gravitational mass) cats indeed behave like fermions.

The quantum behavior of cats was, in fact, already noted in the literature. In his celebrated (but ultimately, cruel) thought experiment, Schr\"odinger of course implicitly assumed that his cat is in a coherent state when it is enclosed in a box that may bring about the poor animal's demise \cite{halpern2015einstein}. More recently, Heinlein \cite{heinlein1985cat} made the explicit observation that cats, especially younger felines that do not yet know that they are not supposed to do so, can quantum tunnel through solid walls\footnote{Similar quantum tunneling behavior was demonstrated by the author's cat Rigby, who successfully tunneled from the bottom to the top of a cat house, and was fortuitously observed while doing so (Fig.~\ref{fig:quantumcat}).}. The seminal work of these authors encourages us to view the coherent quantum behavior of cats as a serious proposal.

A fermionic representation of cats also leads to the observation that the feline field may have a nonzero macroscopic spin tensor, i.e., a corresponding stress-energy-momentum tensor that is nonsymmetric. However, as we indicated in the preceding section, the gravitational field may be indifferent to any antisymmetric contribution to the stress-energy-momentum tensor due to a nonzero spin density.

The quantum nature of cats may be tested experimentally. In addition to quantum tunneling, noted by Heinlein, it may be possible to observe feline species oscillations. The dynamical behavior of cats may be measurable. Observing feline dynamics, e.g., felines falling in a gravitational field, may allow for precision reconstruction of their stress-energy-momentum tensor, explicitly demonstrating its lack of symmetry.

Finally, we note that the assumed massless nature of the ``cheshire cat'' state is reminiscent of the possibility that one of the three neutrino mass states may in fact be a massless state. Systematic study of feline behavior in a gravitational field may offer further insight and take us closer to a robust determination of neutrino masses.

\section{Discussion}
\label{sec:summary}

In this study, we discussed the {\em passive} gravitational mass as being manipulated by felines. This was done on purpose, in an abundance of caution. While it is possible, perhaps even likely, that passive and active gravitational mass are really the same, to measure the latter would require measuring the gravitational field of a cat in motion. Such fields are much too weak, and subject to far too much unpredictable noise to be measurable in practice.

Assuming that the behavior we observed applies to the entire family of {\em felidae}, perhaps in the future it will be possible to carry out similar experiments with a large cat species, such as a lion or tiger. Considerably more massive than {\em felis catus}, perhaps the gravitational field of these large cats is independently observable, confirming the equivalence of passive and active gravitational mass. However, such experiments are obviously not without danger, and unfortunately, the present author neither has access to such large felines for experimental purposes, nor has the requisite training to deal with such animals safely.

Although we assumed that the observed quantity is the gravitational mass of the animal, what is actually measured is mass times Newton's gravitational coupling parameter $G$, i.e., $Gm$, not $m$. In Einstein's field equations, more precisely the Einstein--Hilbert Lagrangian, the quantity $G$ determines the strength of the gravitational field. Cats do not appear to influence $G$, as it would have an impact on the gravitational behavior of other, nearby objects, an effect that we do not see. However, it is perhaps possible that a future modified theory of gravitation introduces instead a {\em bona fide} coupling term, one that appears on the ``matter'' side of the Einstein--Hilbert Lagrangian. In that case, another physical interpretation becomes possible: Perhaps what cats manipulate is not their gravitational mass but the coupling between their mass and the gravitational field. This would also offer a natural explanation as to why the observed behavior affects only the cats' gravitational, not inertial, mass.

As we have seen, the fascinating dynamical behavior of cats, along with its apparent quantum mechanical nature opens up several possible avenues for further research. It is our intention to continue our investigation. Results, when obtained, will be reported elsewhere.


\section*{Acknowledgments}

\begin{quotation}
\noindent{\small\setstretch{0.9}\myinit  thanks Valer Dragan for discussions, and acknowledges the contributions of Freddy, Raina and Rigby for showcasing their abilities of gravity manipulation. Contributions of Marzipan, on account of his successful demonstration of feline levitation, are also postfelidously recognized.\par}
\end{quotation}

\vskip -1.55em\asterism

\begin{quotation}
\noindent{\small\setstretch{0.9}\myinit  has no conflicts to disclose. No cats were harmed, experimented upon, or otherwise inconvenienced in the creation of this study. All feline participants were simply observed in their natural, gravity-defying antics, with treats provided as compensation when appropriate.\par}
\end{quotation}

\bibliography{cats}

\end{document}